\documentclass{article} % For LaTeX2e
\usepackage[preprint]{colm2026_conference}

\usepackage{microtype}
\usepackage{hyperref}
\usepackage{url}
\usepackage{booktabs}
\usepackage[pdftex]{graphicx}
\usepackage{lineno}
\usepackage{amsmath}
\usepackage{fontawesome}
\usepackage{graphicx}
\usepackage{caption}
\usepackage[most]{tcolorbox} % nice boxes (optional but recommended)
\usepackage{xcolor}

\definecolor{darkblue}{rgb}{0, 0, 0.5}
\definecolor{grey}{rgb}{0.65,0.65,0.65} 
\hypersetup{colorlinks=true, citecolor=darkblue, linkcolor=darkblue, urlcolor=darkblue}
\usepackage{newunicodechar}
\newunicodechar{−}{\textminus}

\title{
\name{}:
Can Vibe Coders Really Pass the Vibe Check?}

\author{
Srijan Bansal$^1$, 
Jiao Fangkai$^{2,3}$, 
Yilun Zhou$^1$\thanks{Work done during Salesforce tenure.}, 
Austin Xu$^1$*, 
Shafiq Joty$^{1,2}$, 
Semih Yavuz$^1$ \\
$^1$ Salesforce AI Research, $^2$ Nanyang Technological University, $^3$ A*STAR\\
}

\newtoggle{comments}
\togglefalse{comments}

\iftoggle{comments}{
    \newcommand{\srijan}[1]{\textcolor{orange}{(Srijan: #1)}}
    \newcommand{\fk}[1]{\textcolor{blue}{(Fangkai: #1)}}
    \newcommand{\yilun}[1]{\textcolor{ForestGreen}{(Yilun: #1)}}
    \newcommand{\austin}[1]{\textcolor{orange}{(Austin: #1)}}
    
    \newcommand{\shafiq}[1]{\textcolor{red}{(Shafiq: #1)}}
}{
    \newcommand{\srijan}[1]{}
    \newcommand{\fk}[1]{}
    \newcommand{\yilun}[1]{}
    \newcommand{\austin}[1]{}
    \newcommand{\shafiq}[1]{}
}

\newcommand{\name}{\textsc{VibePass}}

\begin{document}

\ifcolmsubmission
\linenumbers
\fi

\maketitle
\begin{abstract}
As Large Language Models shift the programming toward human-guided ''vibe coding'', agentic coding tools increasingly rely on models to self-diagnose and repair their own subtle faults---a capability central to autonomous software engineering yet never systematically evaluated. 
% We present the first empirical decomposition of \emph{fault-targeted reasoning} across 12 frontier LLMs, structured around two coupled tasks: \emph{Fault-Triggering Test Generation (FT-Test)} , constructing a discriminative witness that exposes a latent fault, and \emph{Fault-targeted Program Repair (FPR)} , repairing it under varying diagnostic conditions.
% (--> Replaced with below) 
We present \name{}, the first empirical decomposition that jointly evaluates two coupled tasks: \emph{Fault-Triggering Test Generation (FT-Test)} constructing a discriminative witness that exposes a latent bug, and \emph{Fault-targeted Program Repair (FPR)}, repairing it under varying diagnostic conditions. 
% To support this analysis, we construct \name{} benchmark from competitive programming problems paired with LLM-generated solutions that pass partial test suites but fail on semantic edge cases---precisely the failure signature of deployed AI coding tools. Unlike prior benchmarks that evaluate test generation and repair in isolation, \name{}  measures the full diagnostic chain, enabling controlled identification of where autonomous debugging breaks down. 
% (--> Replaced with below) 
\name{} pairs competitive programming problems with LLM-generated solutions that pass partial test suites but fail on semantic edge cases, enabling controlled identification of where the diagnostic chain breaks down. 

% We find that fault-targeted reasoning does not scale with general coding ability: models produce syntactically valid tests at near-ceiling rates yet collapse on discriminative generation, with fault hypothesis generation---not output validation---as the dominant bottleneck. 
% (--> Replaced with below) 
Evaluating 12 frontier LLMs, we find that fault-targeted reasoning does not scale with general coding ability. Models produce syntactically valid test inputs at near-ceiling rates yet collapse on discriminative generation, with fault hypothesis generation---not output validation---as the dominant bottleneck.
% When self-generated tests do successfully witness a fault, the resulting repair matches or outperforms repair guided by externally provided tests, revealing that contextual alignment matters more than test quality alone. 
% (--> Replaced with below) 
Test-guided repair reveals a complementary insight: when self-generated tests successfully witness a fault, the resulting repair matches or outperforms repair guided by externally provided tests, but tests that fail to witness the fault actively degrade repair below unguided baselines. %revealing that contextual alignment matters more than test quality alone. 
% FT-Test failures propagate directly into FPR: models that cannot localize a fault cannot repair it, and tests that fail to witness the fault actively degrade repair performance. 
% (--> Removed this and blended partially with the next)
% Together, these results establish fault targeted reasoning as the binding bottleneck in autonomous debugging---a capability that remains deeply deficient across all frontier models.
% (--> Replaced with below) 
Together, these results reframe the challenge of autonomous debugging: the binding bottleneck is not code synthesis or test validity but fault-target reasoning, a capability that remains deficient across all frontier models.

\end{abstract}

\begin{center}
    {\normalsize \faGithub~ \href{https://github.com/SalesforceAIResearch/vibepass}{github.com/SalesforceAIResearch/vibepass}}~\\
    {\normalsize \includegraphics[height=11pt]{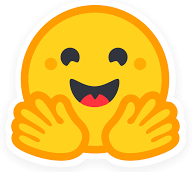}~ \href{https://huggingface.co/datasets/Salesforce/vibepass}{huggingface.co/datasets/Salesforce/vibepass}}\\
    
\end{center}

     % While frontier models excel in isolated coding tasks their joint success rate on individual problem instances collapses. Gemini-3-Pro and GPT-5.2 maintain the most stability, dropping from 0.67–0.88 on single skills to 0.40–0.45 joint agreement. In contrast, Anthropic models exhibits a severe performance gap, falling from $\sim0.9$ in isolated evaluation to near-zero (0.05–0.07) joint success. This ``agreement collapse'' proves that high benchmark scores mask a fundamental inability to apply diverse reasoning skills consistently to the same coding problem.

\section{Introduction}
Recent advances in Large Language Models (LLMs) have accelerated \emph{vibe coding}—agentic workflows that generate substantial software with minimal human oversight~\citep{dong2025survey}. While recent systems surpass 90\% pass rates on established benchmarks~\citep{austin2021mbpp,livecodebench,chen2021humaneval}, this performance reflects generation under ideal conditions: producing correct solutions from clear specifications. Real-world deployment demands more—models must verify code, diagnose latent faults not captured by existing tests, and synthesize targeted repairs. This gap represents a core failure mode of production coding assistants ~\citep{evalplus,trickybugs,liu2025llmpoweredtestcasegeneration}. In practice, LLM-generated code satisfies all visible tests yet fails silently on boundary inputs. This raises a foundational question: \textbf{given a near-correct program with no observable failures, can an LLM synthesize a concrete input witnessing the latent fault and exploit that diagnosis to repair it?} 

We term this \emph{fault-targeted reasoning}. Existing work evaluates its components in isolation: test generation research~\citep{Altmayer_Pizzorno_2025,chen2024chatunitestframeworkllmbasedtest} uses syntactic validity and branch coverage as \emph{proxies} for fault exposure, while repair research~\citep{tian-etal-2024-debugbench,chopra2024exploringinteractionpatternsdebugging} assumes a \emph{declared} fault location, short-circuiting diagnosis entirely. Neither measures the connective capability binding autonomous debugging. We show that proxy metrics overestimate fault-detection capability, and that this omitted diagnostic step is the binding bottleneck for end-to-end reliability.

\begin{figure*}[t]
    \centering
    \includegraphics[width=0.8\linewidth]{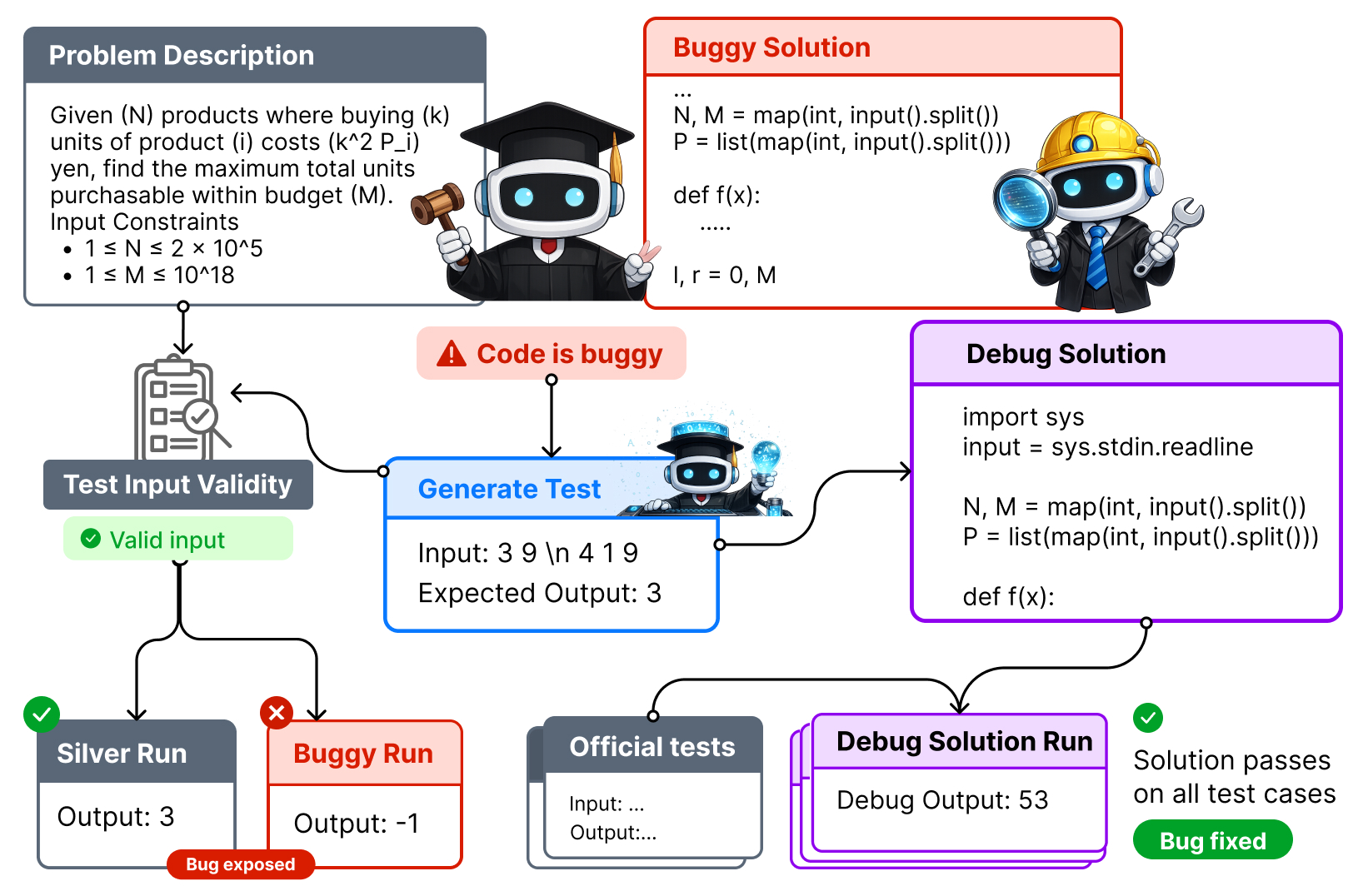}
    \caption{\textbf{\name{}} evaluates LLM performance across roles requiring fault-targeted reasoning, as typical in practical coding agents. Given a problem description and a buggy solution, the LLM (\textit{Judge}) first determines whether a bug exists. If a bug is detected, the LLM (\textit{Tester}) generates a fault-triggering (FT) test, consisting of an input and expected output \textbf{(FT-Test Bug Discovery)}. An FT-test is correct if it satisfies three conditions: the input is valid, the buggy solution fails the test, and a silver solution passes it. The FT-test is then used by the LLM (\textit{Debugger}) to produce a revised solution \textbf{(Fault-Targeted Program Repair)}, which must pass the official test suite to be considered a valid fix.}
    % \shafiq{Why not use the same test from the CCG in the right ((test) optional)}} 
    \label{fig:pipeline}
\end{figure*}

We introduce \textbf{\name{}}, a benchmark that systematically evaluates fault-targeted reasoning through two complementary tasks (Figure~\ref{fig:pipeline}). \textbf{Fault-Triggering Test Generation (Task-1)} measures whether LLMs can generate tests that expose bugs in model-generated solutions—code that passes trivial test cases but contains subtle semantic flaws. Unlike existing test generation benchmarks that focus on coverage or correctness verification, we require tests to be \emph{discriminative}: valid inputs on which buggy and accepted implementations diverge. \textbf{Fault-Targeted Program Repair (Task-2)} evaluates the ability to fix these bugs under three guidance conditions—no tests, externally provided fault-triggering tests, and self-generated tests—enabling controlled investigation of how diagnostic context influences repair success. \name{} consists of 173 instances spanning 76 algorithmically challenging problems from LiveCodeBench, each paired with a model-generated buggy solution and a platform-accepted human-authored solution. Buggy solutions pass 10–90\% of official test cases, ensuring bugs are non-trivial and require semantic reasoning to detect. 

We study fault-targeted reasoning as a \emph{raw reasoning capability} across 12 frontier models—including GPT-5 variants, Gemini-3, Claude Opus/Sonnet-4.6, and open-source systems—revealing three critical findings. First, fault-targeted reasoning does not scale with general coding ability: while 86\% of inputs are syntactically valid, only 61\% are fault-triggering (ranging from 80\% to 26\%), with \emph{fault hypothesis generation} as the dominant bottleneck (2.7× larger gap than output validation). Second, self-generated tests match or outperform external ones—improving repair by 6.4 points for strong reasoners when both yield valid corner cases —indicating that \emph{contextual alignment} matters more than test quality alone. Third, fault-triggering input generation and output validation are near-perfectly coupled ($r=0.98$), strongly predicting repair success ($r=0.79$). We identify performance cliffs of multi-stage repair pipelines at fault localization (−15 pp) and test-to-repair transition ($\sim$21 pp) confirming that fault-targeted reasoning—not code or test synthesis—is the binding bottleneck. Our contributions are: \textbf{(i)} a multi-stage framework decoupling test generation from repair; \textbf{(ii)} a high-quality benchmark with execution-based verification and multi-setting evaluation; and \textbf{(iii)} the first systematic analysis across 12 frontier LLMs, revealing substantial deficits in causal program reasoning despite strong generation performance. \name{} benchmark data and evaluation code are publicly available.

\section{\name{}}
\label{sec:benchmark}
\name{} introduces a benchmark for fault-targeted reasoning through two complementary tasks (Fig~\ref{fig:pipeline}): \textbf{(1) Fault-triggering Test Generation (FT-Test)}, which evaluates the ability to craft fault-triggering (FT) tests that expose subtle semantic bugs, and \textbf{(2) Fault-targeted Program Repair (FPR)}, which measures the ability to repair these bugs under varying guidance. \name{} focuses on model-generated buggy solutions---code that passes trivial tests but contains semantic flaws, reflecting how LLM bugs grow increasingly nuanced as model capability improves. Our design follows three core principles: diverse semantic bugs, execution-based verification, and multi-setting evaluation (Appendix~\ref{app:desgin}).

\subsection{Benchmark Construction}
\label{sec:benchmark-construction}

We construct \name{} via a three-stage pipeline to ensure high-quality, verifiable instances that emphasize non-trivial reasoning (details in Appendix.~\ref{app:construction}).

\noindent\textbf{Problem Collection.} We focus on function-level algorithmic problems, which enable precise execution-based evaluation and fine-grained test generation. We source  problems from LiveCodeBench~\citep{livecodebench}, whose continuous release mitigates train-test contamination. After discarding problems with conflicting test cases or non-standard evaluation, we apply capability-based filtering, retaining only problems unsolved by majority of the three reasoning models (GPT-4o, Claude Sonnet 4, Gemini-2.5-Pro), yielding 170 problems, 89\% rated ``medium'' or ``hard''.

\noindent\textbf{Input-Validity Checker Generation.}
Evaluating test generation requires verifying that generated inputs satisfy 
problem constraints. For each problem, we automatically generate a Python checker 
$\texttt{is\_valid}(t_i) \rightarrow \{\text{True},\text{False}\}$ using 
\texttt{gpt-5-mini}, validated against official test cases to ensure no 
false negatives---yielding valid checkers for 98.8\% of problems (See Figure \ref{fig:prompt_output} for sample prompt input and model-generated output)

\noindent\textbf{Solution Collection, Verification and Filtering.}
We collect 2,184 candidate solutions across 168 problems
from human submissions and diverse LLMs (Appendix~\ref{app:solutions}). Solutions 
passing all official tests are designated \textbf{silver} \footnote{Accepted solutions passing all test cases are considered \emph{silver} because the provided test suites may be incomplete; passing them does not guarantee absolute correctness.}; wrong-answer 
failures are designated \textbf{buggy}---runtime errors are discarded as 
trivially detectable. We retain one silver and at most 4 buggy solutions per 
problem, prioritized first by difficulty then by semantic diversity

\noindent\textbf{Final Dataset.}
\name{} contains 173 instances spanning 76 unique problems. Each instance comprises a problem specification, official test cases, an input-validity checker, verified human-authored silver solution, and a model-generated buggy solution. Bug difficulty ranges from 10\%--90\% pass 
rate on official tests. Table~\ref{tab:dataset_stats} provides 
full statistics.

\subsection{Task 1: Fault-Triggering Test Generation (FT-Test)}
\label{sec:task1}

\textbf{Motivation.} Effective fault localization requires targeted tests that expose bugs~\citep{rafi2024futurestudyingdatacleanness, rafi2025sbestspectrumbasedfaultlocalization}---central to test-driven development, and automated program repair. Task 1 evaluates whether LLMs can generate \emph{FT tests}: inputs that trigger buggy behavior while passing correct implementations.

\textbf{Task Definition.}
Given a problem specification $P$ and buggy solution $S_{\text{buggy}}$, generate a test 
$(t_i, t_o)$ where $t_i$ indicates valid input and $t_o$ is the expected output. A test 
is a \emph{FT test} if: $S_{\text{buggy}}(t_i) \neq t_o \land S_{\text{silver}}(t_i) = t_o$.

\textbf{Evaluation Settings.} We evaluate two settings reflecting different information conditions: In \textbf{Bug-Aware FT-Test}, the model is informed that $S$ contains a bug and must generate a test exposing it---measuring reasoning about known faults. In \textbf{Bug-Discovery FT-Test}, the model must first determine whether $S$ is 
correct or buggy, then generate a FT-test only if it judges $S$ 
buggy---measuring combined fault detection and test generation. Together, 
these settings isolate the impact of bug awareness on fault-targeted 
reasoning.

\textbf{Evaluation Metrics.} We evaluate \texttt{FT-Test} task along four progressively stringent criteria:
\begin{itemize}
    \item \textbf{Validity ($V_{I}$):} $t_i$ conforms to the input specification, verified by the checker.

    \item \textbf{Executability ($V_{IO}$):} $t_i$ is valid and 
    $S_{\text{silver}}(t_i) = t_o$.

    \item \textbf{Discriminative Input ($D_I$):} $t_i$ is valid and 
    $S_{\text{buggy}}(t_i) \neq S_{\text{silver}}(t_i)$.

    \item \textbf{Discriminative Test ($D_{IO}$):} $(t_i, t_o)$ satisfies 
    executability s.t. $S_{\text{buggy}}(t_i) \neq t_o$.
\end{itemize}

Each level strictly refines the previous: $D_{IO} \subseteq \{V_{IO}, D_I\} 
\subseteq V_I$. Critically, the $(V_I \!\to\! D_I)$ gap captures failures in 
fault hypothesis generation while the $(V_I \!\to\! V_{IO})$ gap captures 
failures in output validation---two distinct failure modes.

We additionally measure \textbf{Judgment Accuracy ($J$)}---whether the model 
correctly classifies solution correctness---and \textbf{$J$+*}, the 
conjunction of correct judgment and the corresponding \texttt{FT-Test} metric (e.g., 
$J\!+\!D_{IO}$), capturing end-to-end success.

\subsection{Task 2: Fault-targeted Program Repair (FPR)}
\label{sec:task2}

\textbf{Motivation.}
Generating tests that expose bugs serves a diagnostic purpose---models must also \emph{repair} identified faults. Task 2 evaluates whether LLMs can fix subtle semantic bugs under varying guidance.

\textbf{Task Definition.} Given a problem specification $P$ and a buggy solution $S_{\text{buggy}}$, generate a corrected solution $S_{\text{fixed}}$ that passes all official test cases.

We evaluate three settings to investigate whether external tests, self-generated tests, or no additional context beyond knowing a bug exists best supports program repair.
\texttt{FPR-NoTest}: the model receives no test cases, establishing a baseline for unguided repair ability.
\texttt{FPR-ExtTest}: the model receives a FT test generated in Task-1 as explicit evidence of faulty behavior, measuring the utility of externally provided diagnostic context.
\texttt{FPR-IntTest}: the model first generates its own fault-targeted test and then uses it to guide repair, measuring whether self-constructed reasoning scaffolds aid debugging.

\textbf{Evaluation Metrics.} The primary metrics are \textbf{Pass@1} and \textbf{Success Rate (SR)}—the percentage of fixes passing all official test cases (partial fixes count as failures), aggregated across problems and instances. For \texttt{FPR-IntTest}, we also assess self-generated test quality using Task 1 metrics ($V_{IO}, D_{IO}$) to examine whether successful debugging aligns with high-quality test generation.

\subsection{Evaluated Models}
\label{sec:models}

We evaluate 12 models spanning frontier commercial and open-source systems: 
\textbf{OpenAI} %\citep{singh2025openaigpt5card} 
(GPT-5-Nano, GPT-5-Mini, GPT-5.2, GPT-5.2-Codex), 
\textbf{Gemini} %\citep{comanici2025gemini25pushingfrontier}
(Gemini-3-Flash, Gemini-3-Pro, Gemini-3.1-Flash-Lite, 
Gemini-3.1-Pro), \textbf{Claude} (Sonnet-4.6, Opus-4.6), and 
\textbf{open-source} including GPT-OSS-120B and Nemotron-3-Nano-30B-A3B.
%~\citep{openai2025gptoss120bgptoss20bmodel}
%~\citep{glm5team2026glm5vibecodingagentic}
 Full experimental details are provided in Appendix~\ref{app:llm}.

\section{Results and Analysis}
\label{sec:results}

We organize our findings around three research questions that probe the capabilities and failure modes of frontier LLMs in fault-targeted reasoning. Each subsection presents quantitative results, identifies key failure patterns, and analyzes statistical dependencies across task components.

\subsection{RQ1: How effectively do frontier LLMs generate fault-triggering test cases?}
\label{sec:rq1}

\begin{table*}[t]
\centering
\caption{\textbf{Bug-Aware and Bug-Discovery FT-Test Generation Performance.} 
In the Bug-Aware, the model knows the code is buggy, and performance measures FT-Test generation quality directly: $V_I$/$V_{IO}$ for input/output validity, and $D_I$/$D_{IO}$ for discriminative effectiveness. 
In the Bug-Discovery, J denotes judgment accuracy (buggy or not), and joint metrics (J+*) require both correct judgment and the corresponding quality criterion. Pass@1 is also reported to compare FT-Test with code generation (CG) reasoning ability.}
\label{tab:ccg_results}
\begingroup
\resizebox{\textwidth}{!}{
\begin{tabular}{@{}lc|cccc|ccccc@{}}
\toprule
\textbf{Model} & \textbf{CG}
& \multicolumn{4}{c|}{\textbf{Bug-Aware FT-Test}} 
& \multicolumn{5}{c}{\textbf{Bug-Discovery FT-Test}} \\
\cmidrule(lr){2-2} \cmidrule(lr){3-6} \cmidrule(lr){7-11}
& $P_{@1}$ & \textbf{$V_I$} & $V_{IO}$ & {$D_I$} & {$D_{IO}$} 
& \textbf{$J$} & J+$V_I$ & J+$V_{IO}$ & J+$D_{I}$ & J+$D_{IO}$ \\
\midrule

Gemini-3.1 Flash-Lite & 25.0 & 89.6 & 69.4 & 31.8 & 26.0 & 39.9 & 33.0 & 24.3 & 20.8 & 15.6 \\
Gemini-3 Flash & 78.9 & 95.4 & 92.5 & 63.6 & 63.0 & 75.1 & 74.0 & 69.4 & 60.1 & 59.0 \\
Gemini-3 Pro & 81.6 & 94.8 & 89.6 & 72.3 & 69.4 & 83.2 & 80.4 & 71.1 & 64.2 & 59.5 \\
Gemini-3.1 Pro & 92.1 & 89.0 & 83.2 & 69.9 & 68.8 & 80.4 & 77.5 & 59.5 & 59.5 & 48.0 \\
\midrule

GPT-5 (nano) & 48.7 & 92.5 & 82.1 & 57.8 & 52.0 & 54.9 & 52.0 & 28.9 & 35.3 & 17.9 \\
GPT-5 (mini) & 43.4 & 89.0 & 82.1 & 63.0 & 60.7 & 62.4 & 60.7 & 54.3 & 48.0 & 45.7 \\
GPT-5.2 & 71.1 & 88.4 & 82.1 & 69.4 & 68.2 & 83.8 & 81.5 & 76.3 & 66.5 & 65.3 \\
GPT-5.2 (codex)& 71.1 & 90.2 & 85.6 & 72.8 & 72.3 & 85.0 & 82.1 & 75.7 & 64.2 & 63.6 \\
\midrule

Sonnet-4.6 & 57.9 & 80.4 & 71.7 & 71.7 & 70.5 & 78.6 & 75.7 & 69.9 & 69.9 & 69.9 \\
Opus-4.6 & 73.7  & 84.4 & 80.4 & 79.8 & 79.8 & 82.7 & 79.8 & 75.7 & 75.1 & 75.1 \\
\midrule

GPT-OSS-120B & 67.1 & 76.3 & 69.9 & 66.5 & 65.3 & 74.0 & 70.5 & 66.5 & 67.1 & 65.9 \\
Nemotron-3-30B-A3B & 46.0 & 66.5 & 49.1 & 47.4 & 39.3 & 57.2 & 52.0 & 15.0 & 39.9 & 13.3 \\
\midrule
\textbf{Mean} & \textbf{63.1} & \textbf{86.4} & \textbf{78.1} & \textbf{63.8} & \textbf{61.3} & \textbf{71.4} & \textbf{68.3} & \textbf{57.2} & \textbf{55.9} & \textbf{49.9} \\

\bottomrule
\end{tabular}}
\endgroup
\end{table*}

Table~\ref{tab:ccg_results} reveals a consistent hierarchy of performance across all evaluated models. In the Bug-Aware \texttt{FT-Test} setting, LLMs generate syntactically valid inputs with high reliability (average $V_I = 86.4$), yet only $61.3$ achieve discriminative FT-tests ($D_{IO}$), yielding a 25.5\% gap between input validity and discriminative FT-test generation. This result extends classical observations from automated testing that test adequacy and fault detection are fundamentally distinct capabilities~\citep{10.1145/3324884.3416667} to the LLM regime. The gap arises because producing format-compliant inputs largely relies on learned specification patterns, whereas exposing faults requires causal reasoning about program behavior. High $D_{IO}$ performance dispersion under identical task conditions (e.g., ~54pp gap between Opus-4.6 (79.8) and Gemini-3.1 Flash-Lite (26.0), underscores that FT-test generation is a highly discriminative capability that does not scale uniformly with general coding proficiency.

Our framework decomposes Task-1 failures into two axes: the fault hypothesis gap ($V_I - D_I$, avg. $\approx 23$ pp) and the output validation gap ($V_I - V_{IO}$, avg. $\approx 8$ pp). The fault hypothesis gap is 2.7x larger, showing that identifying FT-test inputs is the dominant bottleneck. Fault hypothesis gap varies 12× across models—from 57.8\% (Gemini-3.1 Flash-Lite)  to 4.6\% (Opus-4.6)—making it a reliable discriminator of capability. A sharp capacity cliff is visible at the lower end: GPT-5-Nano achieves 92.5\% $V_I$ but collapses to 57.8\% $D_I$, and Gemini-3.1-Flash-Lite falls to 31.8\% $D_I$ despite 89.6\% $V_I$. Moreover, $D_{IO}/D_I \approx 0.96$ versus $V_{IO}/V_I \approx 0.9$, indicating that models that locate discriminating inputs also predict outputs accurately, so the output validation gap is largely subsumed by the fault hypothesis bottleneck. 

In Bug-Discovery, models must first judge solution correctness, creating a two-stage bottleneck. Average judgment ($J$) is 71.4\%, with end-to-end $J+D_{IO}$ falling to 49.9\% versus 61.3\% in Bug-Aware $D_{IO}$, and 60\% of failures trace to misjudgment. Judgment dominates weaker models, while conditional test generation differentiates stronger ones—e.g., Gemini-3.1 Pro drops 32.4 pp from $J$ to $J+D_{IO}$, versus ~7.6 for Opus 4.6. Remaining failures split into a conditional fault hypothesis gap ($\sim$12 pp) and output validation gap ($\sim$11 pp).
  
To isolate degradation induced by the judgment requirement, we compare the FT-Test (bug-discovery vs. bug-aware) settings conditioned on judgment in the bug-discovery setting (Table~\ref{tab:t1_v_t2_model_performance}). Bug-aware outperforms bug-discovery for 6 of 12 models, confirming that models effectively leverage self-derived judgment—making external grounding redundant or even detrimental. When the model correctly identifies a bug (J=1), bug-aware outperforms bug-discovery for only 6 of 12 models, suggesting that many models effectively leverage self-derived judgments without requiring external grounding. For these models (e.g., Opus-4.6, Sonnet-4.6, GPT-5.2-Codex), providing the bug label offers minimal or even negative value---the model's internal confidence already guides effective test generation. The clearest exception is GPT-5-Nano ($J = 54.9\%$), where external bug grounding yields substantial gains ($D_{IO}|J$: 66.3\% vs.\ $D_{IO}|{\sim}J$: 34.6\%). This asymmetry indicates that weaker models benefit from explicit grounding because their internal judgment is unreliable. Conversely, for stronger models, the bug-aware setting can depress performance by forcing test generation even in low-confidence cases---eliminating the implicit abstention mechanism available in bug-discovery. This aligns with the \emph{selective prediction} framework in calibrated LLM evaluation~\citep{wen2025knowlimitssurveyabstention}, where allowing models to abstain on uncertain instances improves aggregate reliability. Moreover, in the bug-aware setting we observe a substantial performance gap conditioned on judgment accuracy: when comparing bug-aware results stratified by bug-discovery judgment correctness, models achieve $D_{IO}|J = 71.2\%$ versus $D_{IO}|{\sim}J = 33.6\%$ on average. This 37.6 pp gap confirms the strong coupling between judgment accuracy and test quality, even when external bug labels are provided---suggesting that the underlying diagnostic capability remains the primary bottleneck rather than the availability of grounding information.

% The clearest exception is GPT-5 Nano ($J = 54.9$), where providing the bug location yields a substantial gain ($D_{IO}/J$: 66.3 vs.\ 32.6), indicating that weaker models benefit from explicit grounding because their internal judgment is unreliable. This asymmetry arises because bug-aware forces generation regardless of confidence, eliminating the implicit abstention available in bug-discovery and depressing $D_{IO}$ for models that would otherwise self-select only high-confidence cases—consistent with the \emph{confident abstention} effect in calibrated LLM evaluations~\citep{wen2025knowlimitssurveyabstention}. Moreover, in the bug-aware setting we observe a large gap between $D_{IO}/J = 72.4$ and $D_{IO}/{\sim}J = 32.8$, confirming the strong coupling between judgment accuracy and test quality.

\begin{table*}[t]
\centering
\caption{Model performance on FT-Test Generation under Bug-Discovery and Bug-Aware settings. $J$ denotes judgment accuracy; $J+D_{IO}/J$ measures joint success in discovery. In Bug-Aware, $D_{IO}$ is split by correct ($D_{IO}/J$) and incorrect ($D_{IO}/\sim J$) verdicts. Agreement between discriminative test outcomes across settings is also reported.}
\label{tab:t1_v_t2_model_performance}
\begingroup
\resizebox{0.9\textwidth}{!}{
\begin{tabular}{lccccc}
\toprule
\textbf{Model} & \multicolumn{2}{c}{\textbf{FT-Test {discovery}}} & \multicolumn{2}{c}{\textbf{FT-Test {aware}}} & \textbf{Agreement} \\
\cmidrule(lr){2-3} \cmidrule(lr){4-5} \cmidrule(lr){6-6}

& $J$ & $J+D_{IO}/J$ & ${D_{IO}} /J$ & \textbf{${D_{IO}} /\sim J$} & $J+D_{IO}  \cap  D_{IO}$\\
\midrule
Gemini-3.1 Flash-Lite & 39.9 & 39.1 & 34.8 & 20.2 & 87.0 \\
Gemini-3 Flash  & 75.1 & 78.5 & 73.8 & 30.2 & 87.7 \\
Gemini-3 Pro  & 83.2 & 71.5 & 75.0 & 41.4 & 84.0 \\
Gemini-3.1 Pro & 80.3 & 59.7 & 73.4 & 50.0 & 73.4 \\
\midrule
GPT-5 (nano) & 54.9 & 32.6 & 66.3 & 34.6 & 57.9 \\
GPT-5 (mini) & 62.4 & 73.1 & 76.9 & 33.8 & 83.3 \\
GPT-5.2 & 83.8 & 77.9 & 74.5 & 35.7 & 92.4 \\
GPT-5.2 (codex) & 85.0 & 74.8 & 78.2 & 38.5 & 95.2 \\
\midrule

Sonnet-4.6 & 78.6 & 89.0 & 82.4 & 27.0 & 91.9 \\
Opus-4.6 & 82.7 & 90.9 & 87.4 & 43.3 & 95.1 \\
\midrule
GPT-OSS-120B & 74.0 & 89.1 & 78.9 & 26.7 & 88.3 \\
Nemotron-3 Nano-30B  & 57.2 & 23.2 & 52.5 & 21.6 & 58.6 \\
\midrule
Mean & 71.4 & 66.6 & 71.2 & 33.6 & 82.9 \\
\bottomrule
\end{tabular}}
\endgroup
\end{table*}

\subsection{RQ2: How effectively do frontier LLMs debug subtle semantic bugs?}

Table~\ref{tab:dbg_results} reports debugging performance under three test-guidance settings: \texttt{NoTest} (unguided, model knows the solution is buggy and sees only the problem), \texttt{ExtTest} (model receives an FT-Test from Task 1, correct or incorrect), and \texttt{IntTest} (model first generates its own test before debugging) %\semih{Keep NoTest, ExtTest, IntTest definitions consistent with Section-2.3}.
Across all three FPR settings, proprietary models substantially outperform open-source counterparts, with Gemini-3 Flash achieving the strongest overall profile (NoTest SR: 70.5, IntTest SR: 56.0, ExtTest SR: 71.0) while Nemotron-3 Nano-30B consistently anchors the bottom. Counterintuitively, internally generated tests degrade repair performance on average — mean P@1 drops ~6.8 points from NoTest (58.6) to IntTest (51.8) — indicating that self-generated test scaffolding introduces more noise than signal for most models. Externally generated tests partially recover performance but still fail to surpass NoTest SR on average (45.9 vs. 47.6), and while they consistently produce higher validity ($V_{IO}$: 75.4 vs. 70.1) and discriminative power ($D_{IO}$: 56.3 vs. 54.7) than internally generated ones, this quality advantage does not uniformly translate to higher SR. The finding that both internal and external test augmentation underperform the no-test baseline on average suggests that generated tests — regardless of source — tend to impose repair constraints that are either misleading or misaligned with the true fault, and that models currently lack the robustness to filter and exploit such imperfect signals effectively. Code generation ability further proves to be a poor proxy for repair: GPT-5 (mini) achieves higher IntTest P@1 (52.9) than CG (43.4), Sonnet-4.6 records the table's highest $D_{IO}$ (74.0) despite modest CG (57.9), and GPT-5.2 and its codex variant remain statistically indistinguishable across all FPR metrics, collectively suggesting that repair performance is governed more by fault localization and test utilization capacity than by raw generative capability.

\begin{table}[t]
\centering
\caption{\textbf{NoTest}, \textbf{IntTest}, and \textbf{ExtTest} Fault-Targeted Program Repair Performance. The first four columns (\textbf{left}) report P@1 under different settings. The remaining columns (\textbf{right}) show Success Rate (SR), $V_{IO}$, and $D_{IO}$ (input/output validity and discriminative effectiveness) for internally generated or externally provided tests. Pass@1 for code generation (CG) is also included for comparison. "No", "Int", and "Ext" correspond to NoTest, IntTest, and ExtTest FPR settings, respectively.}
\label{tab:dbg_results}
\begingroup
\resizebox{\textwidth}{!}{
\begin{tabular}{@{}lcccc|ccccccc}
\toprule

\textbf{Model} & \textbf{CG} & \textbf{No} & \textbf{Int} & \textbf{Ext} 
& \textbf{NoTest}  & \multicolumn{3}{c}{\textbf{IntTest}}  & \multicolumn{3}{c}{\textbf{ExtTest}} \\

\cmidrule(lr){2-5} \cmidrule(lr){6-6} \cmidrule(lr){7-9} \cmidrule(lr){10-12}

& \multicolumn{4}{c}{\textbf{$P_{@1}$}} & \textbf{$SR$} & $V_{IO}$ & $D_{IO}$ & \textbf{$SR$} & $V_{IO}$ & $D_{IO}$ & \textbf{$SR$} \\
\midrule

Gemini-3.1 Flash-Lite & 25.0 & 45.7 & 23.4 & 48.1 & 29.0 & \textcolor{grey}{70.5} & \textcolor{grey}{31.5} & 15.5 & \textcolor{grey}{67.5} & \textcolor{grey}{22.5} & 29.5 \\

Gemini-3 Flash & 78.9 & 86.1 & 66.8 & 82.5 & 70.5 & \textcolor{grey}{79.5} & \textcolor{grey}{62.0} & 56.0 & \textcolor{grey}{85.0} & \textcolor{grey}{55.5} & 71.0 \\
Gemini-3 Pro & 81.6  & 80.4 & 64.5 & 78.6 & 68.5 & \textcolor{grey}{86.5} & \textcolor{grey}{70.1} & 53.5 & \textcolor{grey}{90.0} & \textcolor{grey}{65.5} & 67.0 \\
\midrule

GPT-5 (nano) & 48.7 & 38.3 & 35.6 & 39.2 & 24.0 & \textcolor{grey}{65.0} & \textcolor{grey}{47.0} & 23.5 & \textcolor{grey}{81.0} & \textcolor{grey}{49.0} & 28.5 \\
GPT-5 (mini) & 43.4 & 45.2 & 52.9 & 51.6 & 35.0 & \textcolor{grey}{60.5} & \textcolor{grey}{47.0} & 37.5 & \textcolor{grey}{78.5} & \textcolor{grey}{58.0} & 39.0 \\
GPT-5.2 & 71.1 & 69.7 & 69.6 & 59.6 & 65.0 & \textcolor{grey}{76.5} & \textcolor{grey}{65.5} & 64.0 & \textcolor{grey}{77.5} & \textcolor{grey}{65.5} & 58.5 \\
GPT-5.2 (codex) & 71.1 & 67.8 & 68.9 & 61.0 & 63.5 & \textcolor{grey}{71.5} & \textcolor{grey}{63.0} & 60.0 & \textcolor{grey}{78.5} & \textcolor{grey}{66.5} & 54.5 \\
\midrule

Sonnet-4.6 & 57.9 & 63.5 & 64.0 & 57.3 & 48.5 & \textcolor{grey}{63.0} & \textcolor{grey}{51.0} & 48.0 & \textcolor{grey}{74.5} & \textcolor{grey}{71.5} & 43.5 \\
Opus-4.6 & 73.7 & 73.9 & 65.7 & 70.3 & 59.0 & \textcolor{grey}{71.0} & \textcolor{grey}{59.0} & 55.0 & \textcolor{grey}{79.5} & \textcolor{grey}{66.5} & 60.0 \\

\midrule

GPT-OSS-120B & 67.1 & 54.5 & 44.2 & 54.3 & 44.5 & \textcolor{grey}{70.5} & \textcolor{grey}{59.5} & 34.5 & \textcolor{grey}{70.5} & \textcolor{grey}{62.5} & 44.5 \\
% Qwen-3-235B & 62.0 & 39.8 & 14.8 & 28.3 & 28.5 & \textcolor{grey}{22.3} & \textcolor{grey}{19.0} & 10.4 & \textcolor{grey}{64.5} & \textcolor{grey}{54.0} & 22.5 \\
Nemotron-3 Nano-30B & 46.0 & 19.9 & 14.1 & 11.9 & 16.0 & \textcolor{grey}{56.5} & \textcolor{grey}{45.5} & 9.5 & \textcolor{grey}{47.0} & \textcolor{grey}{36.0} & 9.0 \\
\midrule
Mean & 60.4  & 58.6  & 51.8 & 55.9  & 47.6 & \textcolor{grey}{70.1} & \textcolor{grey}{54.7} & 41.6 & \textcolor{grey}{75.4} & \textcolor{grey}{56.3} & 45.9 \\

% rows go here
\bottomrule
\end{tabular}
}
\endgroup

\end{table}
\begin{figure}
    \centering
    \includegraphics[width=\textwidth]{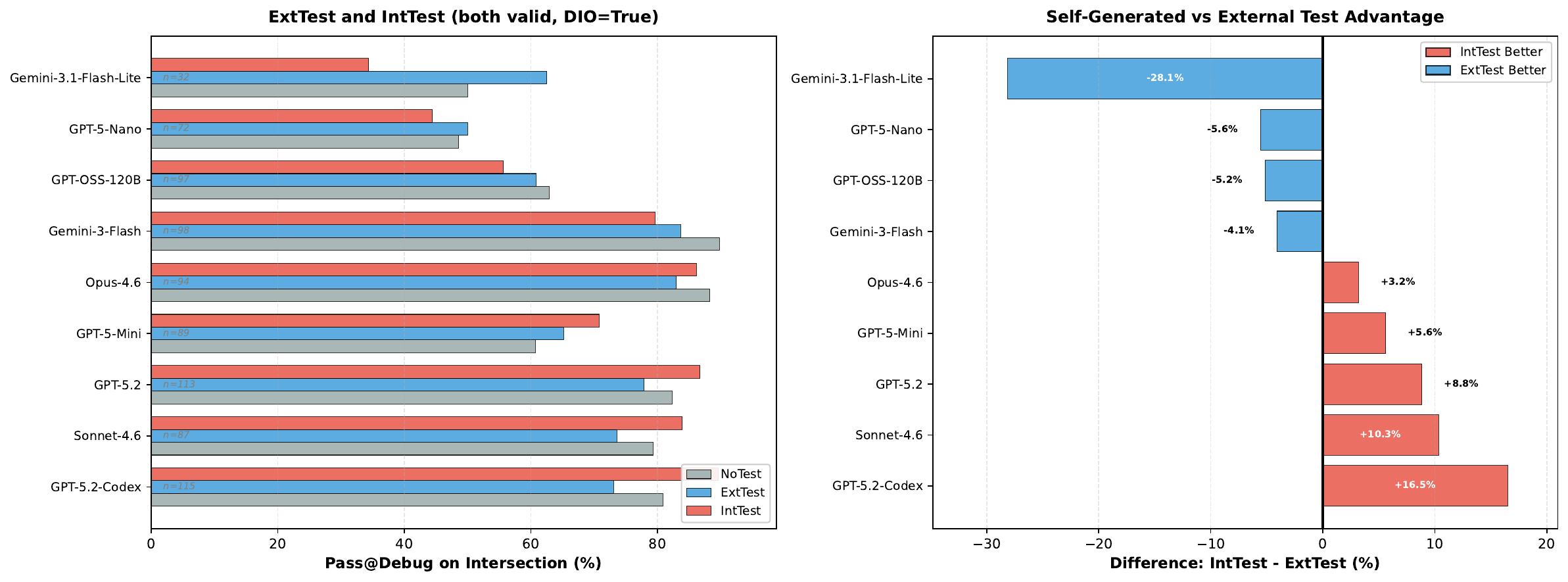}
    \caption{\textbf{Controlled comparison of test feedback mechanisms on valid corner-case intersection}. Debugging performance is shown for samples where both external (Task-1) and self-generated (Task-3) tests are valid. \textbf{Left}: Success rates under \textsc{NoTest} (gray), \textsc{ExtTest} (blue), and \textsc{IntTest} (red) across 11 models. \textbf{Right}: Performance difference (\textsc{IntTest} − \textsc{ExtTest}) per model, with red bars favoring self-generated tests and blue bars favoring external tests.}
    \label{fig:controlled}
\end{figure}

 % We conducted a controlled comparison to isolate the effect of test feedback mechanisms by analyzing samples where both external (Task-1) and self-generated (Task-3) tests produced valid corner-cases, ensuring differences reflect feedback effectiveness rather than test quality. For each model, we evaluated debugging success on the intersection of valid samples (32–115 per model, median 94) under three conditions: no feedback (\textsc{NoTest}), external test
 %  (\textsc{ExtTest}), and self-generated test (\textsc{IntTest}). Across models, \textsc{NoTest} achieved 71.4\%
 %  success, \textsc{ExtTest} 70.0\% (−1.5pp), and \textsc{IntTest} 70.1\% (−1.3pp), with \textsc{IntTest} outperforming \textsc{ExtTest} by 0.2pp on average. At the model level, 5/9 models benefited more from self-generated tests, with the strongest advantage for GPT-5.2-Codex (+16.5pp), while 4/9 models favored external tests, most notably Gemini-3.1-Flash-Lite (−28.1pp), indicating high model-dependent variance. Notably, only 3/9 models showed improvement with external tests over no feedback, and 4/9 with self-generated tests, suggesting that even valid tests can be counterproductive depending on model capacity and architecture. These results suggest that test provenance matters: self-generated tests aligned with debugging context provide marginally more effective guidance than external tests for capable models, highlighting the value of self-diagnosis even when both test types are valid corner-cases.

We conducted a controlled comparison to isolate the effect of test feedback mechanisms by analyzing samples where both external (Task-1) and self-generated (Task-3) tests produced valid corner-cases, ensuring differences reflect feedback effectiveness rather than test quality. For each model, we evaluated debugging success on the intersection of valid samples (32–115 per model, median 89) under three conditions: no feedback (\textsc{NoTest}), external test (\textsc{ExtTest}), and self-generated test (\textsc{IntTest}). Across 12 models, \textsc{NoTest} achieved 63.9\% success, \textsc{ExtTest} 57.8\% (−6.1pp), and \textsc{IntTest} 64.2\% (+0.3pp), with \textsc{IntTest} outperforming \textsc{ExtTest} by 6.4pp (Wilcoxon $p<0.05$, Cohen's $d=0.41$). At the model level, 8/12 models benefited more from self-generated tests, though some favored external tests, indicating high model-dependent variance ($\sigma=20.1\%$). These results suggest that test provenance matters: self-generated tests aligned with the debugging context provide more effective guidance than external tests, highlighting the value of self-diagnosis even when both test types are valid corner-cases.

\begin{figure}[t]
    \begin{center}
    \includegraphics[width=\linewidth]{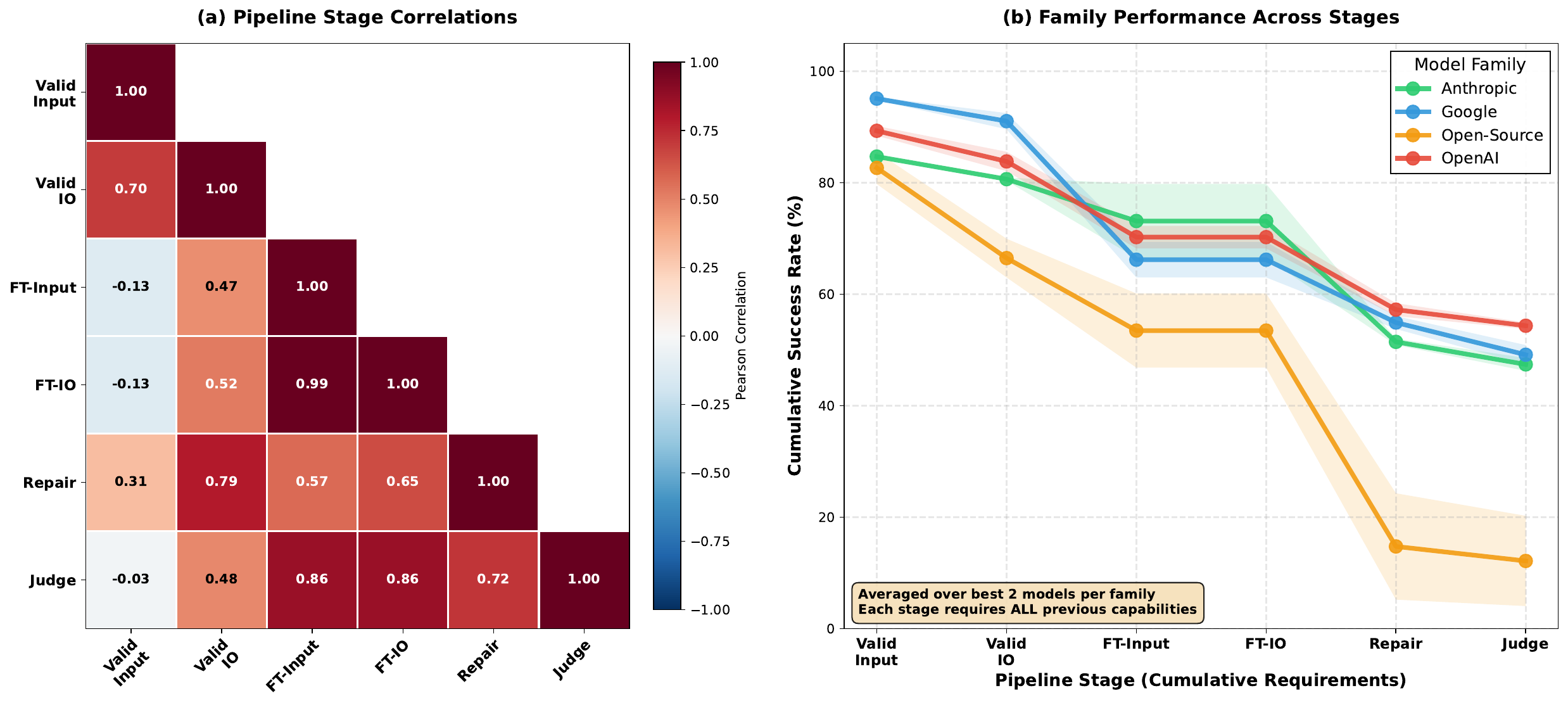}
    \end{center}
    \vspace{-1em}
    \caption{\textbf{Pipeline Stage Correlations and Model Family Performance Across Cumulative Requirements.} Using \name{} instances, we evaluate 12 frontier LLMs across a progression of coding-reasoning tasks: input generation (\textit{Valid Input}), output prediction (\textit{Valid-IO}), fault-triggering discrimination (\textit{FT-Input}, \textit{FT-IO}), and program \textit{Repair}, with \textit{Judge} metrics assessing final correctness. \textbf{[Left]} Pearson correlations reveal that fault-triggering metrics (\textit{FT-Input}--\textit{FT-IO}, $r=0.988$) and their relationship to \textit{Judge} performance ($r \geq 0.86$) are the strongest predictors of success, while \textit{Valid Input} alone weakly predicts downstream results ($r=0.311$). This suggests that the ability to generate fault-revealing tests—rather than mere syntactic validity—is more closely aligned with solving complex bugs. \textbf{[Right]} Cumulative success rates show the largest performance drops at the \textit{Valid IO} $\to$ \textit{FT-Input} (14.7 pp) and \textit{FT-IO} $\to$ \textit{Repair} (21.2 pp) transitions, identifying these as the primary reasoning bottlenecks. Family-level trends highlight that while OpenAI models maintain the highest stability (54.3\%), Google models struggle with fault-triggering tasks and open-source models underperform significantly in repair (12.1\%). \name{} maps the full spectrum from basic test generation to advanced debugging, exposing critical gaps in current model capabilities.}
    \label{fig:main_fig}
\end{figure}

\subsection{RQ3: How correlated are code-reasoning skills, and what limits bug repair?}
% \shafiq{It's better to have a separate RQ for this.}

To understand capability interactions within \name{}, we performed a correlation analysis across six cumulative pipeline stages—\textit{Valid Input} ($VI$), \textit{Valid IO} ($V_{IO}$), \textit{FT-Input} ($D_I$), \textit{FT-IO} ($D_{IO}$), \textit{Repair}, and \textit{Judge}—across all the evaluation models.  Figure \ref{fig:main_fig}a shows the average of top two models per family with variance bands %\semih{what is min-max variance bands?}.
The analysis reveals strong inter-dependencies between fault-triggering capabilities and downstream reasoning. FT-Input and FT-IO exhibit near-perfect correlation (r=0.988), indicating that finding a fault-triggering input is the primary hurdle—producing valid output given one is trivial. These metrics strongly predict both Judge ($r\geq0.86$) and Repair performance($r=0.794$), confirming that generating high-quality FT-tests is foundational to autonomous debugging. In contrast, Valid Input shows weak or negative correlations with all downstream stages ($r\leq0.125$), confirming that syntactic fluency does not transfer to higher-order reasoning.

These dependencies manifest as distinct performance cliffs (Figure~\ref{fig:main_fig}b). The sharpest attrition occurs at \textit{Valid IO} $\rightarrow$ \textit{FT-Input} (−14.7pp) and \textit{FT-IO} $\rightarrow$ \textit{Repair} (−21.2pp), marking the transitions from general execution to fault localization and code modification. OpenAI models achieve the highest final success rate (54.3\%) with the most graceful degradation; open-source models lag substantially, especially in repair (12.1\%). Together, these results pinpoint the fault-targeted reasoning as the critical bottleneck for autonomous debugging.

\section{Related Works}

Early benchmarks for code generation focused on short, function-level programming tasks such as HumanEval and MBPP~\citep{chen2021humaneval,austin2021mbpp}, but have since been largely saturated by code-specialized LLMs~\citep{ds-coder,ds-coder-v2,qwen25coder,codeqwen1.5}. Evaluation efforts have since diverged into competition-level algorithmic reasoning and software engineering benchmarks.

\textbf{Competition-Level Code Generation.}
Benchmarks such as AlphaCode~\citep{alphacode}, APPS~\citep{apps}, and xCodeEval~\citep{xcodeeval} evaluate models on competitive programming tasks. Due to potential contamination from web-scale pretraining, newer benchmarks like LiveCodeBench and LiveCodeBench Pro update problems periodically to better assess generalization~\citep{livecodebench,livecodebench-pro}. EvoCodeBench further reduces leakage by evolving datasets aligned with real repositories~\citep{li2024evocodebenchevolvingcodegeneration}.

\textbf{Software Engineering Code Generation.} Unlike competitive problems, software engineering tasks require long-context understanding and maintainable implementations. \citet{in-coder} highlights the mismatch between standard autoregressive objectives and real editing patterns in InCoder. Benchmarks for realistic workflows include repository-level completion, bug fixing, and full repository synthesis~\citep{repo-bench,jimenez2024swebench,nl2repo}. CodeChain proposes modular self-revision to improve correctness on complex generation tasks~\citep{le2024codechainmodularcodegeneration}.

\textbf{Automatic Debugging, Testing, and Repair.} Debugging and repair involve bug localization, test generation, and patch synthesis. LLMs can leverage execution feedback for self‑correction and iterative refinement~\citep{chen2024teaching,wang-etal-2024-intervenor,pmlr-v202-ni23b,chen2023teachinglargelanguagemodels}. Iterative frameworks such as CodeChain and hierarchical debuggers further improve multi-step reasoning. Real-world repair benchmarks like SWE-bench evaluate fixing genuine GitHub issues with reproducible tests~\citep{jimenez2024swebench}. Recent work trains agents over software evolution histories (SWE-RL) \citep{wei2025swerladvancingllmreasoning} and studies candidate patch ranking (SweRank, ICLR 2026) \citep{reddy2025sweranksoftwareissuelocalization}. Complementary efforts explore unit test synthesis~\citep{CodaMosa} and vulnerability repair benchmarks such as CVE-Bench~\citep{zhu2025cvebenchbenchmarkaiagents}, extending automated program repair evaluation beyond traditional settings.

Despite these advances, most benchmarks still emphasize generation from scratch~\citep{evalplus,livecodebench}, whereas real development involves identifying faults, generating tests to expose them, and repairing underlying issues %\semih{SWE-bench also involves in those, but cited as part of "most benchmarks"}
. These capabilities—collectively termed \emph{fault-targeted reasoning}—remain underexplored, motivating dedicated benchmarks for assessing debugging, testing, and repair in realistic coding workflows.

\vspace{-2mm}
\section{Conclusion}

We introduced \textbf{\name{}}, a benchmark for evaluating fault-targeted reasoning in LLMs—the ability to expose latent bugs through discriminative test generation and perform targeted repair. Across 173 instances and 12 frontier models, we find that fault-targeted reasoning does not scale with general coding ability: the dominant bottleneck is fault hypothesis generation, not code or test synthesis. Self-generated tests match or exceed external ones, highlighting the importance of contextual alignment over raw test quality. Performance cliffs at fault localization and the test-to-repair transition confirm that causal program reasoning remains a critical unsolved capability, even for state-of-the-art systems. We hope \name{} serves as a rigorous foundation for advancing autonomous debugging in real-world deployment settings.

\bibliography{colm2026_conference}
\bibliographystyle{colm2026_conference}

\appendix
\clearpage
\section{Appendix}

\subsection{Benchmark Design \& Statistics}

\subsubsection{Key Design Principles}
\label{app:desgin}
\name{} is built on three core design principles that distinguish it from existing benchmarks:

\noindent\textbf{Diverse Semantic Bugs.} We curate non-trivial buggy solutions that pass a subset of official tests. Bugs with higher pass rates and diverse behaviors (measured via Hamming distance) are prioritized, unlike trivially detectable syntax errors. These bugs capture genuine reasoning failures in model-generated code.

\noindent\textbf{Execution-Based Verification.} All metrics rely on automated execution against official test suite. A solution $S$ passes if it terminates to the expected output on every test $t$ of the official test suite $T$ , i.e., $S(t_i) = t_o\ \forall\, t \in T$. A \emph{fault-triggering test} is a test pair $t = (t_i, t_o)$ where the buggy solution fails while the silver (or accepted) solution succeeds: $S_{\text{buggy}}(t_i) \neq t_o = S_{\text{silver}}(t_i)$. 

\noindent\textbf{Multi-Setting Evaluation.} We evaluate models under varying information conditions (bug-aware vs. bug-discovery FT-Test, unguided (NoTest) vs. test-guided (ExtTest or IntTest) program repair) to isolate different reasoning capabilities and measure how external guidance affects performance.

\subsection{Benchmark Construction Details}
\label{app:construction}

\noindent\textbf{Problem Collection.}
We source 1{,}055 function-level algorithmic problems from LiveCodeBench~\citep{livecodebench}, spanning LeetCode, AtCoder, and Codeforces (May 2023–April 2025). Each problem includes a natural language specification, official test cases with verified outputs, and platform-assigned difficulty labels. We discard problems with conflicting test cases (same input, different outputs) or non-standard evaluation (interactive judges, approximate outputs), ensuring deterministic execution-based metrics. We apply \textit{capability-based filtering}, excluding problems solved within 3 attempts by cost-effective frontier models (\texttt{gpt-4o}, \texttt{gemini-2.5-pro}, \texttt{claude-sonnet-4}), retaining only problems requiring substantive reasoning. This yields 170 problems (16\% retention), of which 89\% are rated ``medium'' or ``hard,'' spanning dynamic programming (28\%), graph algorithms (22\%), data structures (18\%), greedy algorithms (15\%), and other domains (17\%).

\noindent\textbf{Input-Validity Checker Generation.}
We prompt \texttt{gpt-5-mini} to produce a Python function 
$\texttt{is\_valid}(t_i) \rightarrow \{\text{True}, \text{False}\}$ that 
checks whether input $t_i$ conforms to all constraints in the problem 
specification. Figure \ref{fig:prompt_output} shows prompt template and model-generated sample. Each checker is validated against the 
official test suite; checkers rejecting any official input are discarded and 
regenerated (up to 3 attempts). This ensures no false negatives, though it 
does not guarantee coverage of all implicit constraints. In practice, 98.8\% 
of problems yield valid checkers, with 100\% acceptance on official tests.

\begin{figure}[t]
\centering

\begin{minipage}[t]{\textwidth}
\begin{tcolorbox}[title=Prompt,colback=gray!5,colframe=black]
\small
Given the following problem statement and starter code, write only the Python function `$is\_valid\_test$` (with the same signature as in the starter code, but without `self`). The function should return True if the inputs are valid according to the problem constraints, and False otherwise. Output only the code for the `$is\_valid\_test$` function, with no extra text or comments.\\

Problem Statement:
You are given a 0-indexed integer array nums. You can perform any number of operations, where each operation involves selecting a subarray of the array and replacing it with the sum of its elements. For example, if the given array is [1,3,5,6] and you select subarray [3,5] the array will convert to [1,8,6].\\Return the maximum length of a non-decreasing array that can be made after applying operations.\\A subarray is a contiguous non-empty sequence of elements within an array.\\ \\Example 1:\\\\Input: nums = [5,2,2]\\Output: 1\\Explanation: This array with length 3 is not non-decreasing.\\We have two ways to make the array length two.\\First, choosing subarray [2,2] converts the array to [5,4].\\Second, choosing subarray [5,2] converts the array to [7,2].\\In these two ways the array is not non-decreasing.\\And if we choose subarray [5,2,2] and replace it with [9] it becomes non-decreasing. \\So the answer is 1.\\\\Example 2:\\\\Input: nums = [1,2,3,4]\\Output: 4\\Explanation: The array is non-decreasing. So the answer is 4.\\ \\Constraints:\\\\$1 <= nums.length <= 10^5$\\$1 <= nums[i] <= 10^5$\\

Starter Code:
\begin{lstlisting}
class Solution:
    def findMaximumLength(self, nums: List[int]) -> int:
        pass
\end{lstlisting}
\end{tcolorbox}
\end{minipage}
\hfill

\begin{minipage}[ht]{0.98\textwidth}
\begin{tcolorbox}[title=Sample Output,colback=gray!5,colframe=black]
\begin{lstlisting}
def is_valid_test(nums: list) -> bool:
    if not isinstance(nums, list):
        return False
    n = len(nums)
    if not isinstance(n, int) or n < 1 or n > 10**5:
        return False
    for x in nums:
        if not isinstance(x, int) or isinstance(x, bool):
            return False
        if x < 1 or x > 10**5:
            return False
    return True
\end{lstlisting}
\end{tcolorbox}
\end{minipage}

\caption{Example prompt and corresponding model output for input-validity generation.}
\label{fig:prompt_output}
\end{figure}

\noindent\textbf{Solution Collection and Verification.}
\label{app:solutions}
We collect candidates from two sources: \textbf{human solutions} (accepted 
submissions scraped from official platforms, serving as high-confidence silver 
solutions) and \textbf{model-generated solutions} (5--10 samples per model per 
problem from diverse LLMs spanning capabilities and architectures; 
Appendix~\ref{app:llm}). All candidates are executed against official test 
suites with relaxed time limits (2$\times$ platform limits) to prioritize 
functional correctness over efficiency. Solutions passing all test cases are 
designated \textbf{silver}; those failing at least one are designated 
\textbf{buggy}. We retain only \emph{wrong-answer} failures, discarding 
runtime errors (exceptions, timeouts), as wrong-answer bugs reflect semantic 
reasoning failures. We retain one human-authored silver solution with minimal execution time per problem and buggy solutions prioritized first by difficulty (higher official pass rate) then by diversity---among solutions with the same pass count, those with maximally different failure patterns measured by Hamming distance over the binary pass/fail vector on official tests. This ensures retained bugs are both subtle (near-correct 
behavior) and semantically distinct. We employ a diverse set of state-of-the-art language models across different capabilities and architectures: (1) OpenAI: GPT-5-Nano, GPT-5-Mini, GPT-5, GPT-5.1, GPT-5.2; (2) Google: Gemini-2.5-Flash-Lite, Gemini-2.5-Flash, Gemini-2.5-Pro, Gemini-3-Flash, Gemini-3-Pro; (3) Anthropic: Claude Opus-4, Opus-4.1, Opus-4.5, Sonnet-4, Sonnet-4.5. This setup enables generation of buggy solutions across families, sizes, and architectures. We manually collected accepted Python solutions from the official coding platform (Leetcode, Atcoder) and verified them against the official test cases. In total, we collect 2{,}184 candidate solutions across 168 problems (13/problem).

\begin{table}[t]
  \centering
  \caption{Dataset statistics for the \name{} benchmark.}
  \label{tab:dataset_stats}
  \small
  \begin{tabular}{lc}
  \toprule
  \textbf{Statistic} & \textbf{Value} \\
  \midrule
  Total instances (problem, gold, buggy tuples) & 173 \\
  Unique problems & 76 \\
  Avg. instances per problem & 2.3 \\
  \midrule
  \multicolumn{2}{l}{\textit{Problem Difficulty (platform labels)}} \\
  \quad Easy & 1\% \\
  \quad Medium & 25\% \\
  \quad Hard & 73\% \\
  \midrule
  \multicolumn{2}{l}{\textit{Buggy Solution Pass Rate on Official Tests}} \\
  \quad 1st quartile & 43\%\\
  \quad Median & 71\%\\
  \quad 3rd quartile & 86\% \\
  \quad Mean & 61\%\\  
  \midrule
  \multicolumn{2}{l}{\textit{Pass Rate Difficulty Labels}} \\
  \quad Near-Correct (80\% or more)  & 31\% \\
  \quad Medium-Correct (40-80\%) & 27\% \\
  \quad Near-Incorrect (40\% or less) & 35\% \\
  \midrule
  \multicolumn{2}{l}{\textit{Problem Domains}} \\
  \quad Array/Matrix & 25\% \\
  \quad Other & 19\% \\
  \quad Dynamic Programming & 18\% \\
  \quad Math/Number Theory & 16\% \\
  \quad Graph Algorithms & 8\% \\
  \quad Other & 13\% \\
  \midrule
  \multicolumn{2}{l}{\textit{Bug Categories}} \\
  \quad Missing Edge Case & 32\% \\
  \quad Incorrect Condition & 28\% \\
  \quad Off-by-One Error & 27\% \\
  \quad Wrong Algorithm & 8\% \\
  \quad Others (DP State Transition Error, Incorrect Initialization) & 5\% \\
  \midrule
  \multicolumn{2}{l}{\textit{Platforms}} \\
  \quad Leetcode & 57\% \\
  \quad Atcoder & 43\% \\
  \bottomrule
  \end{tabular}
  \end{table}

\subsection{Experimental Setup-- LLM Generators for Bug and Solution Generations}
\label{app:llm}
\textbf{Models} We evaluate a diverse set of frontier and open-source language models spanning different capability levels and architectures. Our evaluation includes: (1) OpenAI models (medium reasoning): GPT-5-Nano, GPT-5-Mini, and GPT-5.2, GPT-5.2-Codex; (2) Google models: Gemini-3.1-Flash-Lite, Gemini-3-Flash, Gemini-3-Pro and Gemini-3.1-Pro; (3) Anthropic models: Claude Opus-4.6 and Sonnet-4.6; and (4) Open-source models: NVIDIA-Nemotron-3-Nano-30B-A3B-BF16 (Nemotron-3-Nano), GPT-OSS-120B (high), and Qwen3-235B-A22B-Thinking-2507. This selection enables comparison across different model families, sizes, and architectural choices. For benchmark construction we emply GPT-5-Mini for constructing input validity checkers and weak-to-moderate reasoners (GPT-4o, Gemini-2.5-Pro, Claude Sonnet-4) for capability-bases filtering.

\textbf{Inference Configuration.} For all test generation and debugging tasks, we set temperature T=1 for API models (if available).We set the maximum context window to 128K tokens for open-source models, while  almost 1M+ for API-based models, sufficient to accommodate problem specifications, code solutions, and generated test cases. We use high reasoning effort across all the models.

\textbf{Prompt Design.} All tasks use structured prompts that include: problem description, input-output format specifications, and task-specific instructions. For bug-discovery CCG, prompts explicitly request both a correctness judgment and a test case (if buggy). For test-guided debugging, failing test cases are provided in the prompt with clear formatting. For self-guided debugging, prompts instruct models to generate test cases internally before attempting repairs. Full prompt templates are provided in the Github repository. %Appendix~\ref{app:prompts}.

\textbf{Evaluation Protocol.} All generated solutions and test cases are executed in isolated sandboxed environments with relaxed time limits (5× the original competitive programming limits) to focus evaluation on semantic correctness rather than efficiency. Test validity is verified using the generated input-validity checkers. We report Pass@1 for debugging tasks and FT-Test accuracy metrics for test generation tasks, computed over the full set of benchmark instances. Each experiment is run once due to computational constraints.

\end{document}